\documentclass[12pt,preprint]{aastex}

\shorttitle{Bulges or Bars from Secular Evolution?}

\shortauthors{Debattista et al.}

\begin{document}

\def\etal{{et al.}}
\def\eg{{\it e.g.}}
\def\etc{{\it etc.}}
\def\ie{{\it i.e.}}
\def\cf{{\it cf.}}
\def\degrees{^\circ}
\def\spose#1{\hbox to 0pt{#1\hss}}
\def\gtsim{\mathrel{\spose{\lower.5ex \hbox{$\mathchar"218$}}
     \raise.4ex\hbox{$\mathchar"13E$}}}
\def\ltsim{\mathrel{\spose{\lower.5ex\hbox{$\mathchar"218$}}
     \raise.4ex\hbox{$\mathchar"13C$}}}
 
\title{Bulges or Bars from Secular Evolution?}

\author{Victor P.~Debattista}
\affil{Institut f\"ur Astronomie, ETH H\"onggerberg, CH-8093,
  Z\"urich, Switzerland}
\email{debattis@phys.ethz.ch}

\author{C. Marcella Carollo}
\affil{Institut f\"ur Astronomie, ETH H\"onggerberg, CH-8093,
  Z\"urich, Switzerland}
\email{marcella.carollo@phys.ethz.ch}

\author{Lucio Mayer}
\affil{Department of Theoretical Physics, University of Z\"urich,
  Winterurestrasse 190, 8057, Z\"urich, Switzerland}
\email{lucio@physik.unizh.ch}

\author{Ben Moore}
\affil{Department of Theoretical Physics, University of Z\"urich,
  Winterurestrasse 190, 8057, Z\"urich, Switzerland}
\email{moore@physik.unizh.ch}

\begin{abstract}
  We use high resolution collisionless $N$-body simulations to study
  the secular evolution of disk galaxies and in particular the final
  properties of disks that suffer a bar and perhaps a bar-buckling
  instability.  Although we find that bars are not destroyed by the
  buckling instability, when we decompose the radial density profiles
  of the secularly-evolved disks into inner S\'ersic and outer
  exponential components, for favorable viewing angles, the resulting
  structural parameters, scaling relations and global kinematics of
  the bar components are in good agreement with those obtained for
  bulges of late-type galaxies. Round bulges may require a different
  formation channel or dissipational processes.
\end{abstract}


\section{Introduction}

Evidence has accumulated in the past decade showing that many bulges,
especially at low-masses, have a disk-like, almost-exponential radial fall-off
of the stellar density \citep{as94,cdjb96,dj96,csm98,csdzsd01,c99,mch03}, and
in some cases disk-like, cold kinematics \citep{k93,kbb02}.  Comparisons of
bulge and disk parameters have furthermore shown a correlation between the
scale-lengths of bulges and disks \citep{dj96,mch03} and, on average, similar
colors in bulges and inner disks \citep{tdfdw94,pb96,cdjb96}.  The disk-like
properties of bulges and the links between bulge and disk properties have been
suggested to indicate that bulges may form through the evolution of disk
dynamical instabilities such as bars, which are present in about $70\%$ of
nearby disk galaxies \citep{k99,efp00}.

It has long been known that bars lead to angular momentum
redistribution and to an associated increase in the central mass
density.  Already \citet{h71} found that an initially single component
disk evolved a double exponential density profile under the influence
of a bar; as a result, the scale-length of the outer disk increases.
It has been suggested that bars may be efficient at building {\it
  three-dimensional} stellar bulge-like structures via scattering of
stars at vertical resonances \citep{cdfp90}, or by the collisionless
buckling instability (which weakens the bar; Raha et al. 1991), or via
bar destruction due to the growth of a central mass concentration
\citep{pn90,nsh96}.  As buckling occurs relatively easily, it could be
a particularly promising avenue for bulge formation.  This instability
leads to the large-scale, coherent bending of the bar perpendicular to
the plane of the disk \citep{rsjk91}.  Buckling, which is a result of
vertical anisotropy \citep{a85,fp84,mh91,ms94}, thickens the stellar
system and weakens the bar.  Evidence that buckling occurs in nature
has relied on the fact that buckled bars are boxy/peanut-shaped when
viewed edge-on and on the unmistakable gas-kinematic signature of a
bar observed in such bulges \citep{km95,mk99,bf99}.  The relevance of
buckling to bulge formation remains however rather anectdotal, as
little work has been done to check that the structural and kinematic
properties of buckled bars are quantitatively consistent with those
observed in bulges.

In this Letter, we report on high resolution $N$-body simulations of
bar-unstable disks, some of which buckle, and compare their
secularly-evolved structural and kinematic properties with those of
bulges in local galaxies. We describe our simulations in \S 2, present
our results in \S 3 and discuss their implications in \S 4.

\section{Methods}
\label{sec:methods}

\subsection{Rigid halo simulations}

Most of the $N$-body simulations reported in this Letter consist of a
live disk inside a rigid halo, which allows large numbers of particles
and high spatial resolution.  The rigid halos were represented by
either a logarithmic potential, $\Phi_L= \frac{v_{\rm h}^2}{2}~
\ln(r^2 + r_{\rm h}^2)$, or a Hernquist potential $\Phi_H =
-\frac{M_{\rm h}}{r+r_{\rm h}}$.  The initially axisymmetric disks
were modeled by the generalized surface density profile $I(r) \propto
\exp[-(r/r_o)^{(1/n)}]$ introduced by \citet{s68}.  This gives a
\citet{dv48} profile for $n=4$ and an exponential profile for $n=1$;
we used $1 \leq n \leq 2.5$.  The disks have scale-length $R_{\rm d}$,
mass $M_{\rm d}$, Gaussian thickness $z_{\rm d}$ and are truncated at
a radius $R_t$.  Disk kinematics were set up using the epicyclic
approximation to give constant Toomre-$Q$, which we varied from $1.2$
to $2.0$.  Vertical equilibrium was obtained by integrating the
vertical Jeans equation.  The disks were represented by $4-7.5\times
10^6$ equal-mass particles.  We use units where $R_{\rm d} = M = G =
1$; thus the unit of time is $(R_{\rm d}^3/GM)^{1/2}$.

The simulations were run using a 3-D cylindrical polar grid code
\citep{sv97} with $N_R\times N_\phi \times N_z = 60 \times 64 \times
243$.  The radial spacing of grid cells increases logarithmically from
the center, reaching to $\sim 2 R_t$; in all cases, $R_t = 5$.  The
vertical spacing, $\delta z$, of the grid planes was set such that
$0.2 \geq z_{\rm d} \geq 0.025 \geq 4 \delta z$.  We used Fourier
terms up to $m=8$ in the potential, with a Plummer softening length
$\epsilon = 0.017$.  We verified that our results are not sensitive to
resolution by running the most strongly buckled model with $m=32$ and
smaller $\delta z$.  Time integration was performed with a leapfrog
integrator with a fixed time-step, $\delta t = 0.01$ for all runs
except for the $n>1$ simulations, for which we used $\delta t =
0.0025$.  We chose values for the halo parameters such that our
rotation curves were approximately flat to large radii.  For the
logarithmic halos, we set $v_{\rm h} = 0.68$, while the Hernquist
halos had $M_{\rm h} = 43.4$.  We measured the amplitudes of the bar,
$A_\phi$, and of buckling, $A_z$, as the normalized amplitudes of the
$m=2$ tangential and vertical density distributions.  We will discuss
the effects of different initial conditions elsewhere.

\subsection{Modeling and measured parameters}

In order to compare our final systems with observed galaxies, we first
obtained, for all simulations, radial density profiles of the disk
mass density distribution at three inclinations, $i=0$ (face-on),
$i=30\degrees$ and $60\degrees$. For $i=0$, we simply computed the
azimuthally averaged mass profile.  For $i=30\degrees$ and
$60\degrees$, we considered 3 orientations of the bar with respect to
the major-axis ($\phi_{\rm bar} = 0, 45\degrees$ and $90\degrees$); we
then measured the 1-D projected surface density profiles with the task
{\sc ellipse} in {\sc iraf}\footnote{{\sc iraf} is distributed by
  NOAO, which is operated by AURA Inc., under contract with the
  National Science Foundation}.

We decomposed these mass profiles into a central S\'ersic component plus
an outer exponential disk; a S\'ersic law is usually used in modeling bulge
light profiles \citep{apb95,g01,mch03}.  We stress that by decomposing the
final density profiles into a ``bulge'' plus disk component we are not
implying that secular evolution has produced three-dimensional bulges.
Nonetheless, for brevity, we will refer to as ``bulges'' the central S\'ersic
components, and as ``disks'' the exponential components.

Our S\'ersic bulge plus exponential disk decompositions are characterized by
five parameters: $\Sigma_{0,d}$, $\Sigma_{0,b}$, the disk and the bulge
central surface densities, respectively; $R_d$, the scale-length of the outer
exponential disk; $R_{b,eff}$, the half-light radius of the bulge; and $n_b$,
the index of the S\'ersic profile.  In light of the ill-conditioned fitting
described by \citet{mch03}, $n_b$ was held fixed with respect to the remaining
four parameters when searching for the best-fitting models; the best-fitting
$n_b$ was then identified as the one giving the smallest $\chi^2$ when
repeating the fits with $0.1 \leq n_b \leq 4$ in steps of 0.1.  The five
parameters derived from the bulge/disk decompositions allow us to compute
three dimensionless structural quantities to compare with observations: $n_b$,
$R_{b,eff}/R_d$ and $B/D$, the ratio of bulge to disk mass (we assumed a
constant mass-to-light ratio when comparing with the light-weighted
measurements available for real galaxies).

We also measured the bulge ellipticity, $\epsilon_{\rm b}$, the line-of-sight
velocity dispersion, $\bar{\sigma}$ (by averaging the corresponding profiles
within some radial range), and $V_p$, the peak line-of-sight velocity within
the same radial range on the disk major-axis. These allow us to compute
$V_p$/$\bar{\sigma}$ to investigate the kinematic properties of the resulting
bulges in the $V_p$/$\bar{\sigma}$ versus $\epsilon_{\rm b}$ plane.  In this
plane, normal bulges are thought to follow the locus traced by isotropic
oblate rotators \citep{bt87}, and bulges that result from the
secular evolution of disks are thought to emerge as systems that are
dynamically colder than isotropic oblate rotators \citep{k93,kbb02}.

\subsection{Live halo simulation}

\begin{figure}[!ht]
\plotone{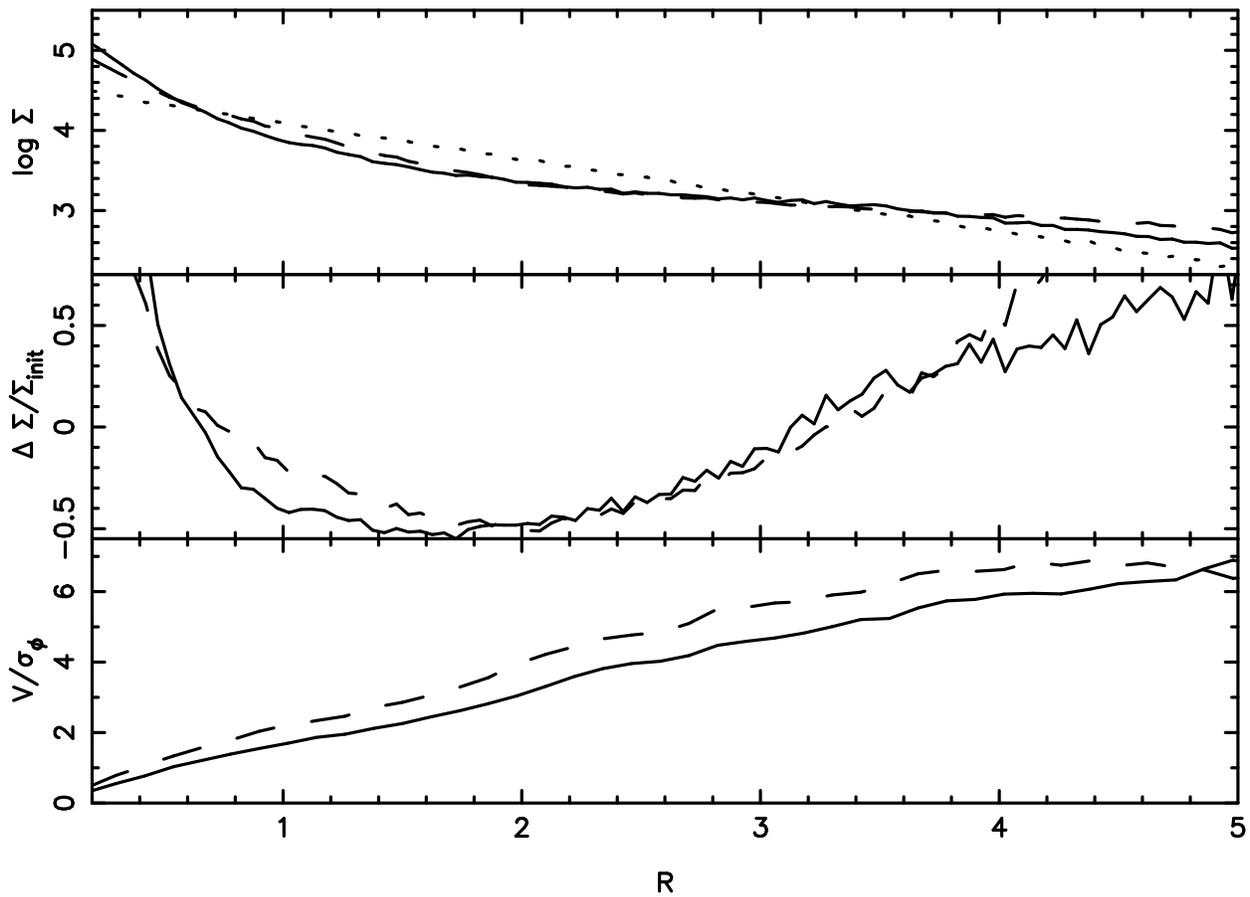}
\caption{
  Comparison of the evolution of the density profiles in matched live (solid
  lines) and rigid halo (dashed lines) simulations after $\sim 2.4$ Gyr.  In
  the top panel, the initial profile is indicated by the dotted line.  The
  fractional change in the profiles, relative to the initial profile, is shown
  in the middle panel.  The bottom panel shows the profile of $V/\sigma_\phi$.
\label{fig:fig1}}
\end{figure}

Rigid halo simulations are better suited to systems in which the disk
is dominant in the inner regions (as are the simulations discussed
here), because the interaction with the halo is then weaker.  While
massive disks represent a reasonable assumption for high surface
brightness barred galaxies \citep{ds00}, it is important to verify
that the full interaction with a live dark matter (DM) halo does not
lead to a drastically different evolution for the structural and
kinematic parameters.
The ability of a bar to lose angular momentum to a DM halo helps make
it stronger \citep{ds00}.  We therefore have run a pair of matched
live and rigid halo simulations.  The live halo simulations were run
on PKDGRAV (\citet{s01}; details of the live halo simulations will be
presented elsewhere).  We checked that these results are not sensitive
to resolution by running the live halo model with larger $N$ and
smaller $\epsilon$.  The two simulations evolved differently: the bar
in the live halo case was slightly stronger and slows down because of
dynamical friction with the halo.  Nonetheless, the mass density and
$V/\sigma_\phi$ profiles in the rigid- and live-halo simulations
remain quite similar, as we illustrate in Fig.  \ref{fig:fig1} (we use
$\sigma_\phi$ in this comparison because it is representative of what
we might expect in high inclination observations).  The larger central
density in the live halo simulation of Fig. \ref{fig:fig1} is a result
of the fact that its bar can shed additional angular momentum.
Otherwise, this comparison suggests that the gross features seen in
the live halo simulation are well-reproduced by the rigid halo
simulation, giving confidence that the rigid halo simulations survey
from which we draw the main conclusions of this paper is adequate to
describe the evolution of massive disks.  Moreover, by using rigid
halos, we assure that we obtain a minimal level of bar secular
evolution, which helps to disentangle the effects of different
evolutionary processes.

\clearpage

\clearpage

\section{Results}
\label{sec:results}

\subsection{Time evolution}

\begin{figure}[!ht]
\plotone{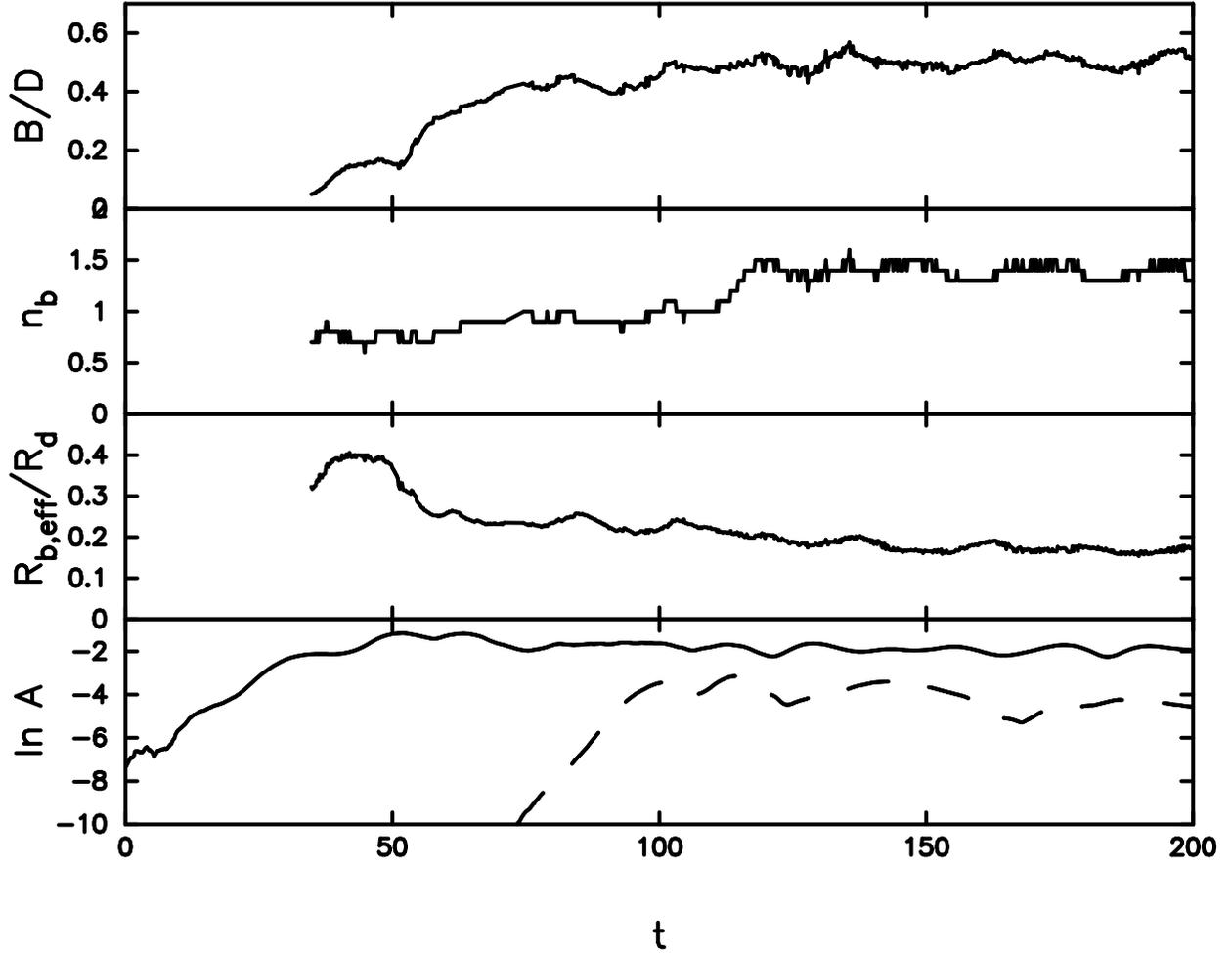}
\caption{
  The evolution of a representative strongly buckling simulation.
  From bottom to top the panels show evolution of $A_\phi$ (strength
  of bar; solid line) and $A_z$ (strength of buckling; dashed line),
  $R_{b,eff}/R_d$, $n_b$ and $B/D$, all measured at $i = 0$.  The bar
  forms at $t\simeq 50$ and buckles at $t\simeq 100$.  Before $t=30$
  the fitting algorithm is unstable and spurious results are obtained
  because the algorithm fits transient spiral structure.
\label{fig:fig2}}
\end{figure}

The evolution of a representative strongly buckling simulation, shown
in Fig.  \ref{fig:fig2}, produced a bar by $t\simeq 50$, which then
buckled strongly at $t\simeq 100$. Despite the strong buckling, the
bar is weakened but not destroyed.  As is well known, the process of
bar formation drives a redistribution of angular momentum, leading to
an increase in the central density.  This process is largely complete
by the time the bar buckles: the buckling instability does not alter
significantly the scale length or the mass of the inner S\'ersic
component.  However, it is interesting that at buckling, $n_b$ changes
from $n_b \simeq 1$ to $n_b \simeq 1.5$: thus buckling may contribute
to the scatter of $n_b$ around the exponential value observed in the
bulges of real intermediate-type galaxies.

\clearpage

\begin{figure}[!ht]
\plotone{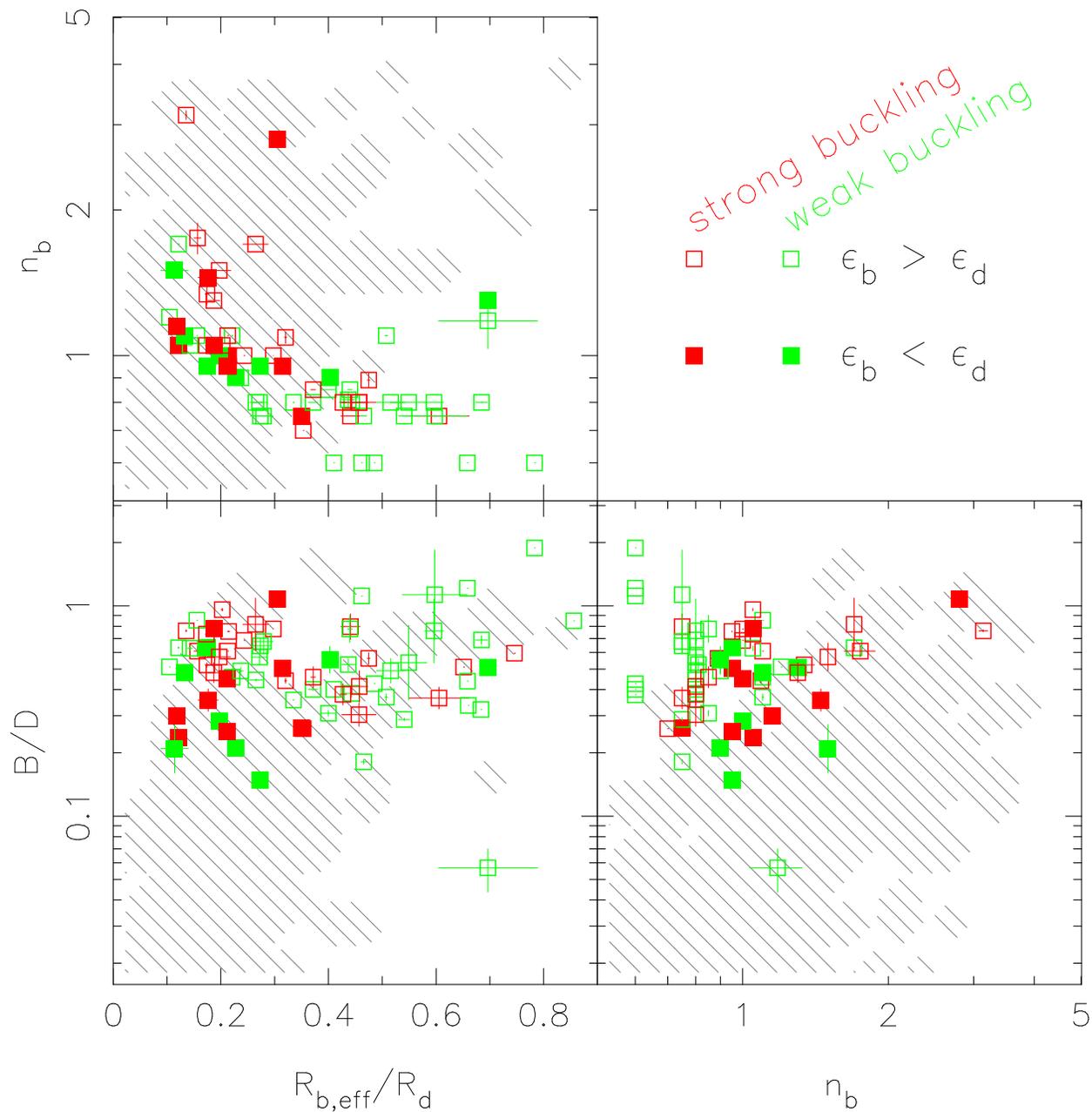}
\caption{
  The structural properties of our simulations compared with observed
  galaxies (shaded area) from \citet{mch03} and \citet{g03}.  Squares
  are filled (empty) when $\epsilon_b < \epsilon_d$ ($\epsilon_b >
  \epsilon_d$) and are red (green) when buckling is strong (weak).
\label{fig:fig3}}
\end{figure}

\clearpage

\subsection{Comparisons with observations}

In Fig. \ref{fig:fig3} we compare the structural properties of our final
systems with similar quantities measured for local bulges.  Our simulations
are represented by filled symbols when $\epsilon_b$ is larger than the
ellipticity of the disk, $\epsilon_d \equiv 1 - \cos i$.  All $i \leq
30\degrees$ projections have $\epsilon_b > \epsilon_d$, as are about half of
the projections at $i= 60\degrees$.  Bulges in our simulations cover a broad
range of parameter space, including regions where no real bulges are found.
This mismatch is diminished when only systems with $\epsilon_b < \epsilon_d$
are considered, but at the cost that a large fraction of the projections,
including all at $i \leq 30\degrees$, are excluded from the sample.
Nevertheless the overlap between the simulations and the observations is quite
good.

\clearpage

\begin{figure}[!ht]
\plotone{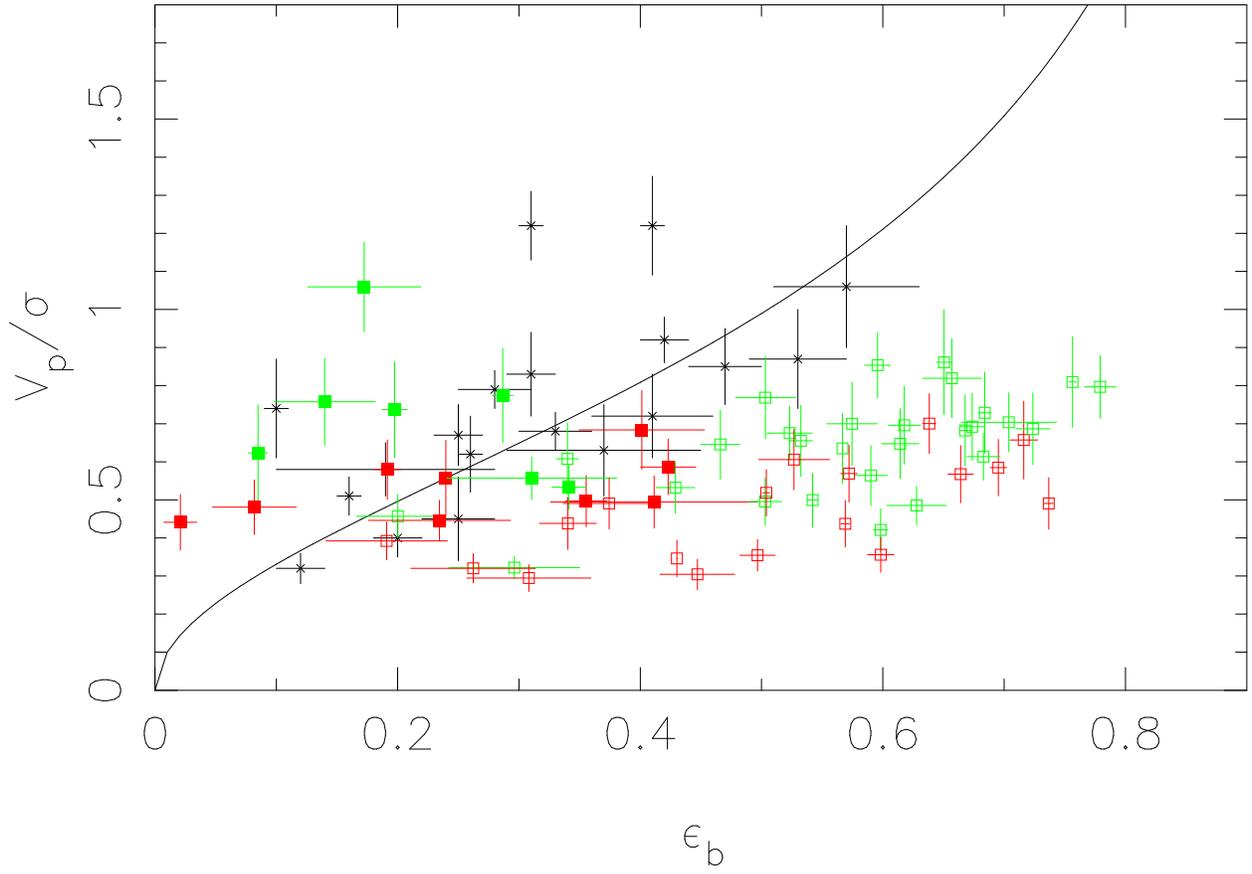}
\caption{
  The $V_p/\bar{\sigma}$ versus $\epsilon_b$ diagnostic plane. Symbols for
  simulations are as in Fig. \ref{fig:fig3} while the observational data of
  \citet{k93} are indicated by black stars.
\label{fig:fig4}}
\end{figure}

\clearpage

We also checked whether global kinematics can distinguish between
bona-fide bulges and bulge-like bars. In Fig.  \ref{fig:fig4} we plot
$V_p/\bar{\sigma}$ versus $\epsilon_b$ for real bulges and for our
simulations.  For real bulges, $V_p$ and $\bar{\sigma}$ were obtained
by fitting a de Vaucouleurs profile to the bulge.  Our simulations are
rather poorly fit by a de Vaucouleurs profile, which introduces a
systematic difference between our measurements and Kormendy's (1993).
To quantify some of this uncertainty, we have measured $\epsilon_b$,
$V_p$ and $\bar{\sigma}$ within $R_{b,eff}$ and $R_{b,eff}/2$ and used
half the difference as our error estimate.
This diagram shows that evolved bars can show, under some viewing
angles, global kinematic-elongation properties typical not only of the
dynamically-cold ``pseudo-bulges'', but also those that are taken as
{\it defining} the bona-fide, dynamically-hot bulges.  These are the
expected projection effects for triaxial spheroids seen at high
inclination \citep{bt87}.  Thus the population of bulges which are
identified by bulge/disk decompositions or by kinematics may have some
contribution from bars.

\section{Discussion}

Although \citet{rsjk91} only reported bar-weakening, not destruction,
by the buckling instability, it has become common wisdom that buckling
destroys bars.  Our simulations have failed to turn up a single
instance in which the bar was destroyed by buckling, although we
cannot exclude that it is in some extreme case.  The channel of bulge
formation by the dissipationless destruction of bars during buckling,
therefore, is not viable.

However, the minimal collisionless secular evolution present in our
simulations must also occur in nature, resulting in systems that exhibit
double component mass density profiles.  For certain viewing orientations, the
spread in structural parameters and kinematic properties are indistinguishable
from those observed in systems that are classified as bulges.

Nonetheless, the existence in nature of round bulges inside low
inclination galaxies, which our simulations cannot reproduce, requires
that other processes are also involved.  Possibly secular evolution
including dissipative gas \citep{mw04} will result in rounder bulges,
but this requires extended central objects with masses of $\sim
10-20\%$ that of the disk \citep{ss04}.  The higher interaction rate
in the early universe may have triggered the large gas inflows needed
to build such objects.

Finally, because the halos of our simulations are rigid, the baryonic
component cannot but conserve its angular momentum.  In (semi-)
analytic models of disk galaxy formation
\citep{fe80,mmw98,ls91,wf91,dss97,vdb98}, the distribution of disk
scale-lengths, $R_d$, is set by that of their angular momenta.
\citet{djl96} found that the width of the observed $R_d$ distribution
at fixed luminosity is smaller than that predicted by analytic theory.
Our simulations show that, under the influence of a bar, $R_d$ may
increase by a factor of 2 or more at constant global angular momentum.
Since less extended disks are likely to be more bar-unstable, the
secular evolution of these disks may be responsible for at least part
of this discrepancy.

\acknowledgments
We would like to thank the anonymous referee for useful suggestions.

\end{document}